\documentclass[paper,published]{JHEP3}
\usepackage{graphicx}
\title{The density of states method at non-zero chemical potential}
\author{
Zoltan Fodor$^{a,b,c}$, Sandor D.~Katz$^b$ and Christian Schmidt$^{a,d}$\\
$^a$Department of Physics, University of Wuppertal, \\
$\;$Gauss 20, D-42119, Germany\\
$^b$Institute for Theoretical Physics, E\"otv\"os University, \\
$\;$P\'azm\'any 1, H-1117 Budapest, Hungary\\
$^c$Department of Physics, University of California at San Diego, \\
$\;$9500 Gilman Drive, La Jolla, CA 92093-0319, U.S.A.\\
$^d$Department of Physics, Brookhaven National Laboratory, \\
$\;$Upton, NY 11973,  U.S.A.\\
$\;$E-mail: \email{fodor@theorie.physik.uni-wuppertal.de, katz@bodri.elte.hu, \\
cschmidt@bnl.gov}
}
\abstract{
We study the QCD phase diagram by first principle lattice calculations at so
far unreached high densities. For this purpose we employ the density of states
method. Unimproved staggered fermions, which describe four quark flavors in the
continuum are used in this analysis. Though the method is quite expensive,
small lattices show an indication for a triple-point connecting three different
phases on the phase diagram.
}
\received{January 25, 2007}
\accepted{March 15, 2007}
\JHEP{03(2007)121}
\begin{document}
\section{Introduction}
To clarify the phase diagram of QCD and thus the nature of matter under
extreme conditions is one of the most interesting and fundamental tasks of
high energy physics. The experimental setup of the running Relativistic Heavy Ion
Collider (RHIC) at the Brookhaven National Laboratory, as well as the
forthcoming Large Hadron Collider (LHC) at CERN, aim on the creation of hot quark
matter, the quark gluon plasma. Future experiments like the Compressed Baryonic
Matter (CBM) experiment at the Facility for Antiproton and Ion Research (FAIR), 
attempt to explore the phase diagram in the region of high temperatures and intermediate
densities. On the theoretical side various color-superconducting phases have been
proposed in the region of very high densities but low temperatures, which may
be relevant to the physics of neutron star interiors. For a review see
for example \cite{Alford}. 

Lattice QCD has been shown to provide important and reliable information from
first principles on QCD at zero density. However, lattice QCD at finite
densities has been harmed by the complex action problem ever since its
inception. At non-zero chemical potentials ($\mu>0$) the determinant of the fermion
matrix ($\rm{det}M$) becomes complex. Standard Monte Carlo techniques using
importance sampling are thus no longer applicable when calculating observables
in the grand canonical ensemble according to the partition function
\begin{equation}
Z_{GC}(\mu)=\int \mathcal{D}U\; \rm{det}M[U](\mu) \exp\{-S_G[U]\}.
\label{eq:Z_GC}
\end{equation}
Note that $Z_{GC}$ is real. The origin of the sign problem are the
fluctuations of the complex phase $\theta$, defined by $\rm{det}M \equiv
\left|\rm{det}M\right|\exp\{i\theta\}$. In case those 
fluctuations ($\sqrt{\left<(\theta -\left<\theta\right>)^2\right>}$) become 
considerably larger than $\pi/2$ \cite{Ejiri:2004yw}, the problem becomes 
serious. For a detailed discussion
of the phase of the fermion determinant see also \cite{Kim}. Recently many different
methods have been developed to circumvent the complex action problem for
$\mu/T\lesssim 1$ \cite{Fodor:2001au, methods}. For a recent 
overview see also \cite{overview}. With all these methods the transition line
of the QCD phase transition was mapped in the ($T$,$\mu$)-plane. The results
coincide for small chemical potentials and the same choice of parameters.

The goal of this paper is to employ the density of states method at non-zero
chemical potential to extend the accessible region of the QCD phase diagram to
lower temperatures and higher densities. The paper is organized as follows: 
In Section~\ref{sec:method} we introduce our method and in
Section~\ref{sec:detail} we give the details of our simulation
parameters. Readers who are not interested in the details of our method or the 
simulation parameters may start at Section~\ref{sec:phases}, where we 
will discussed results on the phase diagram. Results for the quark
number density will be given in Section~\ref{sec:thermodynamics}. Finally we
will conclude in Section~\ref{sec:conclusion}. 

\section{Formulation of the method \label{sec:method}}
A very general formulation of the DOS
method is the following: One special parameter ($\phi$) is held fixed. The
expectation value of a thermodynamic observable ($O$), according to the usual
grand canonical partition function (\ref{eq:Z_GC}), can be recovered by the
integral over $\phi$,
\begin{equation}
<O>=\int d\phi \, \left<Of(U)\right>_\phi \rho(\phi)
\left/ \int d\phi \, \left<f(U)\right>_\phi \rho(\phi)\right.
\label{eq:dos_obs}
\end{equation}
where the density of states ($\rho$) is given by the constrained partition
function:
\begin{equation}
\rho(x)\equiv Z_\phi(x)=\int \mathcal{D}U\, g(U) \, \delta( \phi - x ).
\label{eq:dos_Z}
\end{equation}
With $\left<~\right>_\phi$ we denote the expectation value with respect to the
constrained partition function. In addition, the product of the weight
functions $f,g$ has to give the correct measure of $Z_{GC}$:
$fg=\rm{det}M\exp\{-S_G\}$. This idea of reordering the partition
functions was used in many different cases
\cite{dos, LUO, Azcoiti:2002vk, Ambjorn-RM}.
The advantages of this additional integration becomes
clear, when choosing $\phi=P$ and $g(U)=1$. Here $P$ denotes the plaquette
expectation value. In this case $\rho(\phi)$ is  
independent of all simulation parameters. Once $\rho(\phi)$, as well as the 
correlation between $\phi$ and the observable $O$ is known, $O$ can be calculated as
a continuous function of the lattice coupling $\beta$. If one has stored
all eigenvalues of the fermion matrix for all configurations, the observable
can also be calculated as a function of quark mass ($m$) and number of
flavors \cite{LUO} ($N_f$). 

In this work we chose
\begin{equation}
\phi=P \qquad \mbox{and} \qquad
g=\left|{\rm det}M \right| \exp\{-S_G\}, \qquad f=\exp\{i\theta\}.
\label{eq:con}
\end{equation}
In other words we constrain the plaquette and perform simulations with measure
$g$. This particular choice for the functions $f,g$ was first introduced in 
ref.~\cite{factori} and is called factorization method. It has been successfully
tested on $\theta$-vacuum like systems \cite{Azcoiti:2002vk} and in random matrix 
theory \cite{Ambjorn-RM}. By this choice, one has a clear separation
of the effects coming from the complex phase of the determinant, and the rest.
As we will see later, this enables us to locate the region of the parameter space
which show the fewest fluctuation of the phase and hence contribute most to the 
partition function.  

In practice, we replace the delta function in eq.~(\ref{eq:dos_Z}) by a
sharply peaked potential \cite{Ambjorn-RM}. The constrained partition function
for fixed values of the plaquette expectation value can then be written as
\begin{equation}
\rho(x) \approx \int {\cal D}U\; g(U) \exp\left\{- V(x)\right\},
\end{equation}
where $V(x)$ is a Gaussian potential with
\begin{equation}
V(x)=\frac{1}{2}\gamma\left(x-P\right)^2.
\end{equation}
We obtain the density of states ($\rho(x)$) by the fluctuations of the actual
plaquette $P$ around the constraint value $x$. The fluctuation dissipation
theorem gives
\begin{equation}
\frac{d}{dx}\ln \rho(x)=<\gamma(x-P)>_x.
\end{equation}
Before performing
the integrals in eq.~(\ref{eq:dos_obs}) we carry out a calculation based
on an ensemble generated at $(\mu_0,\beta_0)$:
\begin{eqnarray}
\label{eq:rew1}
\left<Of(U)\right>_x(\mu,\beta)
&=&\left<Of(U)R(\mu,\mu_0,\beta,\beta_0)\right>_x
/\left<R(\mu,\mu_0,\beta,\beta_0)\right>_x,\\
\label{eq:rew2}
\left<f(U)\right>_x(\mu,\beta)
&=&\left<f(U)R(\mu,\mu_0,\beta,\beta_0)\right>_x
/\left<R(\mu,\mu_0,\beta,\beta_0)\right>_x,\\
\label{eq:rew3}
\frac{d}{dx}\ln\rho(x,\mu,\beta)
&=&\left<\gamma(x-P)R(\mu,\mu_0,\beta,\beta_0)\right>_x.
\end{eqnarray}
Here $R$ is given by the ratio of the measure $g$
at the point $(\mu,\beta)$ and at the simulation point $(\mu_0,\beta_0)$,
\begin{equation}
R(\mu,\mu_0,\beta,\beta_0)=g(\mu,\beta)/g(\mu_0,\beta_0)
=\frac{|{\rm det}(\mu)|}{|{\rm det}(\mu_0)|}\exp\{-S_G(\beta)+S_G(\beta_0)\}.
\end{equation}
Having calculated the expressions~(\ref{eq:rew1})-(\ref{eq:rew3}), we are
able to extrapolate the expectation value of the observable~(\ref{eq:dos_obs})
to any point $(\mu,\beta)$ in a small region around the simulation point
$(\mu_0,\beta_0)$. For any evaluation of $\left<O\right>(\mu,\beta)$, we
numerically perform the integrals in eq.~(\ref{eq:dos_obs}). We also
combine the data from several simulation points as described by Eqs.~(\ref{eq:rew1})-(\ref{eq:rew3}).

\subsection{Simulations with constrained Plaquette}
The quantity we want to constrain is the real part of the plaquette
$P_{\mu\nu}(y)$ averaged over lattice points ($y$) and directions ($\mu, \nu$)
\begin{equation}
P=\sum_y\sum_{1\le\mu<\nu\le4} \frac{1}{3} {\rm Re}{\rm Tr}P_{\mu\nu}(y).
\end{equation}
Since the plaquette is also the main part of the gauge action,
\begin{equation}
S_G=-\beta\sum_x\sum_{1\le\mu<\nu\le4}
\left\{\frac{1}{3}{\rm Re}{\rm Tr}P_{\mu\nu}(x)-1\right\} ,
\end{equation}
the additional potential $V$, which constraints the plaquette around a given
value, can be easily introduced in the hybrid Monte
Carlo update procedure of the hybrid-R algorithm \cite{Gottlieb:mq}.
To do so we modify the force in the molecular dynamical evolution of the gauge
field by a factor $(1+\gamma(x-P)/\beta)$. This requires the measurement of
the plaquette in each molecular dynamical step.

\subsection{Generating Configurations with measure $|{\rm det}M(\mu)|$}
For the generation of Configurations with measure $g$, according to
eq.~(\ref{eq:con}) we use the method of ref.~\cite{Kogut:2002tm}. Since at
finite iso-spin chemical potential the fermion determinant is real, our
fermion matrix in flavor space is given by
\begin{equation}
M=\left(\begin{array}{cc} 
M_{KS}(\mu)        & \lambda\gamma_5 \\ 
-\lambda\gamma_5  & M_{KS}(-\mu)             
\end{array}\right).
\label{fermion_matrix}
\end{equation}
Here each component represents a usual
staggered fermion field $\chi$ with four flavors in the continuum limit. The
diagonal elements are thus the usual staggered fermion matrices, in the upper
left corner with chemical potential $\mu$ and in the lower right corner with
chemical potential $-\mu$. The off-diagonal elements are iso-spin symmetry
breaking terms, proportional to the small parameter $\lambda$, which are
necessary in order to see spontaneous symmetry braking on a finite lattice. The
fermion matrix (\ref{fermion_matrix}) will thus represent eight continuum
flavor. Note that the $\gamma_5$ matrices corresponds in the staggered case to
a multiplication of the phase $\epsilon(x)=-1^{x_1+x_2+x_3+x_4}$.  

In order to simulate this system, we use the HMC in complete
analogy to the even-odd ordering of the $\mu=0$ case. This means that we
generate Gaussian noise ($R$) for both components of
$(R_1,R_2)=M(\chi_1,\chi_2)$, but keep only the upper component ($\chi_1)$
after the inversion. This way we still describe eight flavors with the block-diagonal
positive definite matrix $M^\dagger M$ which we use for the evolution of the 
gauge field. In the molecular dynamical integration of
the equations of motion, we perform the usual square-root trick to reduce the
number of flavors to four. In the limit of $\lambda\to 0$ it is however correct
to take the square-root and will not introduce any approximations or locality
problems, since our basic building block is the four flavor staggered fermion
matrix which comes out with a power of one. A further reduction of $n_f$ to two
or one, by using a fractional power of the staggered fermion matrix, introduces
additional difficulties that are still under debate \cite{Sharpe:2006re}. It is
known \cite{trick} that using the square (or forth) root of the staggered matrix
at $\mu>0$ could lead to phase ambiguities . 

\TABLE{
\begin{tabular}{|c|c|c|c|c|c|c|c|c|} \hline
$N_s$ & $N_t$  & $\beta$ & $am$ & $a\mu$ & $P$ & $\lambda$ & \#   \\ \hline \hline
4 & 4 & 4.850 & 0.05 &0.30 & 2.26-3.59 & 0.01, 0.02 &$2\times 44\times 4000$ \\
4 & 4 & 4.850 & 0.05 &0.35 & 2.26-3.59 & 0.01 &$1\times 44\times 4000$ \\
4 & 4 & 4.850 & 0.05 &0.40 & 2.26-3.59 & 0.01, 0.02 &$2\times 44\times 4000$ \\
4 & 4 & 4.850 & 0.05 &0.45 & 2.26-3.59 & 0.01 &$1\times 44\times 4000$ \\
4 & 4 & 4.850 & 0.05 &0.50 & 2.26-3.59 & 0.01 &$1\times 44\times 4000$ \\ \hline
4 & 4 & 4.950 & 0.05 &0.40 & 2.26-2.59 & 0.01 &$1\times 44\times 4000$ \\
4 & 4 & 4.925 & 0.05 &0.40 & 2.26-2.59 & 0.01 &$1\times 44\times 4000$ \\
4 & 4 & 4.900 & 0.05 &0.40 & 2.26-2.59 & 0.01 &$1\times 44\times 4000$ \\
4 & 4 & 4.875 & 0.05 &0.40 & 2.26-2.59 & 0.01 &$1\times 44\times 4000$ \\ \hline 
6 & 6 & 5.10  & 0.05 &0.30 & 2.78-2.98 & 0.01 &$1\times 16\times 5715$ \\
6 & 6 & 5.10  & 0.05 &0.35 & 2.75-2.94 & 0.01 &$1\times 16\times 8987$ \\
6 & 6 & 5.10  & 0.05 &0.40 & 2.72-2.91 & 0.01 &$1\times 16\times 6800$ \\
6 & 6 & 5.10  & 0.05 &0.45 & 2.70-2.89 & 0.01 &$1\times 16\times 2303$ \\ \hline
6 & 6 & 5.18  & 0.05 &0.35 & 2.70-3.23 & 0.01 &$1\times 44\times 3995$ \\
6 & 6 & 5.16  & 0.05 &0.35 & 2.70-3.27 & 0.01 &$1\times 32\times 4393$ \\
6 & 6 & 5.14  & 0.05 &0.35 & 2.70-3.23 & 0.01 &$1\times 32\times 5915$ \\
6 & 6 & 5.12  & 0.05 &0.35 & 2.70-3.05 & 0.01 &$1\times 32\times 6960$ \\ \hline
6 & 8 & 5.10  & 0.05 &0.30 & 2.70-3.16 & 0.01 &$1\times 39\times 2355$ \\ \hline
6 & 8 & 5.10  & 0.03 &0.30 & 2.70-3.14 & 0.006 &$1\times 32\times 1360$ \\ \hline
\end{tabular}
\caption{Simulation parameter and statistics. The number of trajectories is
  given in the last column, here the first factor is due to the number of the
  lambda parameters and the second factor gives the number of plaquette
  values. \label{tab:sim_points}\vspace*{-3mm}}
}
\section{Simulation details and the strength of the sign problem \label{sec:detail}}
Simulations have been performed with staggered fermions and $N_f=4$. We chose
9 different points in the $(\beta,a\mu)$-plane for the $4^4$ lattice and 8
points for the $6^4$ lattice. On each of these points we did simulations with
20-40 constrained plaquette values, all with quark mass $am=0.05$. Further
simulations have been done with $(\beta,a\mu)=(5.1,0.3)$ on the $6^3\times8$
lattice for $am=0.05$ and $am=0.03$. The simulation points and statistics are
summarized in table~\ref{tab:sim_points}.
For each point ($\beta,a\mu$) we have chosen several plaquette values in the
range where the density of states ($\rho$) is large. In addition we measure
all eigenvalues of the reduced fermion matrix. Here ``reduced'' means that the
$\mu$ dependence was reduced by Gauss elimination to the first and last time
slice. This procedure was described in detail in
ref.~\cite{Fodor:2001au}. Having stored the eigenvalues, we are able to
evaluate the fermionic determinant (absolute value and phase) as a function
$\mu$, as well as all their derivatives.

In order to calculate the plaquette expectation value, or its susceptibility,
one has to perform the following integrals:
\begin{equation}
\left<P\right>=\int dx\; x \rho(x) \left<\cos(\theta)\right>_x , \qquad
\left<P^2\right>=\int dx\; x^2 \rho(x) \left<\cos(\theta)\right>_x.
\label{eq:plaq}
\end{equation}
Thus the functions $\rho(x)$ and $\left<\cos(\theta)\right>_x$ have to be known
quite precisely. We plot both functions in figure~\ref{fig:dis}.
\FIGURE{
\label{fig:dis}
\includegraphics[height=7.5cm]{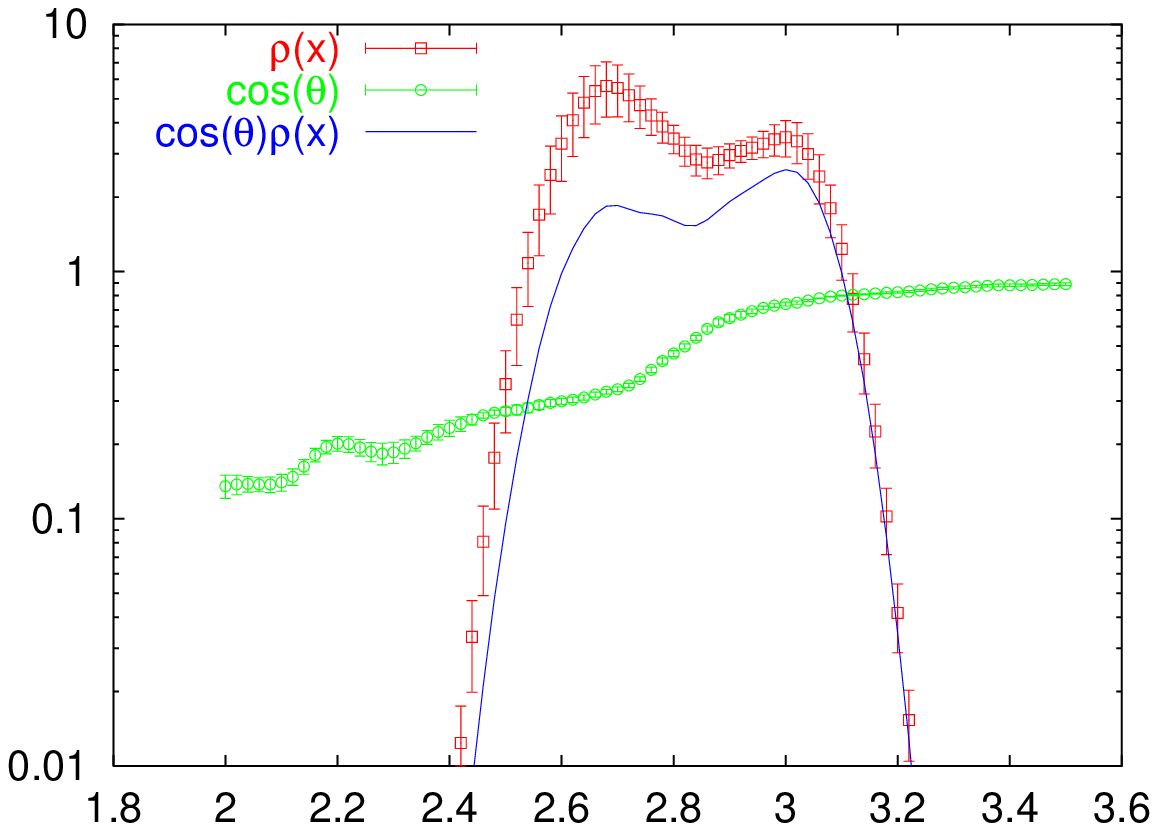}
\caption{Results for simulations at $\beta=4.98$, $\mu=0.3$, $\lambda=0.02$,
  $n_f=4$, $am=0.05$, and number of lattice points: $4^4$. Shown are the density of
  states $\rho(x)$, the phase factor $\left<\cos(\theta)\right>$, and their
  product. }
} 
\FIGURE{
\label{fig:det_phase}
\includegraphics[height=7.5cm]{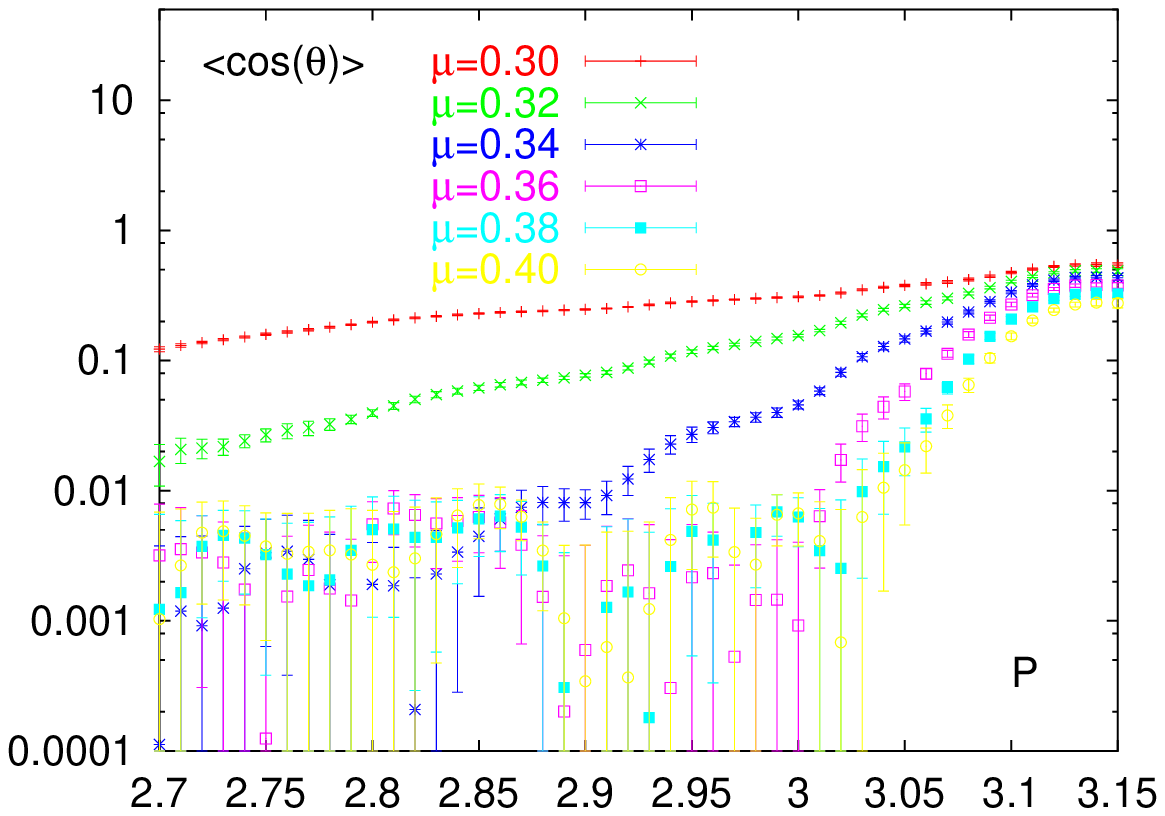}
\caption{Results for simulations at $\beta=5.1$, $\lambda=0.01$,
  $n_f=4$, $am=0.05$, and number of lattice points: $6^4$. Shown is the suppression
  from the complex phase of the fermion determinant $\left<cos(\theta)\right>$ for 
  different chemical potentials. }
}
The transition is signaled in the double peak structure of $\rho(x)$. The phase
factor $\left<\cos(\theta)\right>_x$ suppresses the peak of  $\rho(x)$ at smaller
plaquette values, which results in a shift of the critical temperature to
smaller values, in comparison with the phase quenched theory. 

In figure~\ref{fig:det_phase} we show the phase factor for different chemical
potentials.  
With increasing chemical potential the phase factor
becomes compatible with zero within errors. In fact, its average value becomes
as low as $\cos(\theta)\sim 0.005$. There exists however a small interval
around $P\sim 2.85$, where the phase factor remains finite. In this way, the
Plaquette expectation values are strongly altered by the phase
factor. figure~\ref{fig:det_phase} 
demonstrates also the advantage of he DOS method over the other approaches of
lattice QCD at finite density. Using the DOS method one is able to do
simulations directly at those Plaquette values which are relevant at finite
density. This results in an even better overlap than that of the
multi-parameter re-weighting approach.

In order to be able to transform the lattice units into physical units, we have
measured the heavy quark potential on a zero temperature lattice: $10^3\times
20$. From the potential we determined the Sommer scale $r_0$ and the string
tension $\sigma$. In Addition we have measured the pion mass and the nucleon
mass. The results are given in 
table~\ref{tab:scale}.  
\TABLE[t]{
\begin{minipage}{10cm}
\begin{center}
\begin{tabular}{|c|c|c|} \hline
$\beta$ & $r_0/a$   &$a^2\sigma$ \\ \hline \hline
4.85    & 1.436(18) & 0.818(5)   \\
4.90    & 1.537(84) & 0.745(12)  \\
5.05    & 1.711(35) & 0.576(9)   \\
5.10    & 1.876(16) & 0.445(7)   \\
5.15    & 2.208(17) & 0.321(3)   \\
5.17    & 2.411(4)  & 0.276(1)   \\ \hline \hline
$\beta$ & $am_{\pi}$ & $am_N$    \\ \hline \hline
4.85   & 0.5413(1)  & 2.40(4)    \\       
4.90   & 0.5447(1)  & 2.26(3)    \\       
5.05   & 0.5613(2)  & 2.21(2)    \\       
5.10   & 0.5715(3)  & 2.17(1)    \\       
5.15   & 0.5892(2)  & 2.03(1)    \\       
5.17   & 0.5982(1)  & 1.93(1)    \\ \hline
\end{tabular}
\end{center}
\end{minipage}
\caption{Results for the Sommer radius ($r_0$), string tension ($\sigma$),
  pion mass ($m_{\pi}$) and nucleon mass ($m_N$) for different $\beta$
  values. The results are for $n_f=4$ and $am=0.05$, measured on a $10^3\times
  20$ lattice.\label{tab:scale}}
}
To transform the results from lattice to physical units in practice, the Sommer
parameter $r_0$ was used and has been fitted with a polynomial of the order
three, to get the lattice spacing $a$ as a continuous function of $\beta$. For
$r_0$ in physical units we have used the MILC result of $r_0=0.476(7)(18)$~fm
\cite{MILC}.
 
First of all we check, whether we can reproduce old results with our new method.
For this purpose we reproduce one point of the phase transition line which has been
calculated by multi-parameter re-weighting from $\mu=0$ configurations on the
$4^4$ lattice \cite{Fodor:2001au}.
Performing the integrations in eq.~(\ref{eq:plaq}) 
numerically, we calculate 
the plaquette expectation value and the plaquette susceptibility
$\chi_P\equiv\left<P^2\right>-\left<P\right>^2$.   At any fixed $\lambda$, we
determine the critical coupling
by the peak position of the susceptibility as shown in figure~\ref{fig:sus}.
We indeed find that including the phase factor does shift the transition to
lower values of the coupling, which also means to lower temperatures. This can
be clearly seen in a shift of the peak of the plaquette susceptibility.

Since the $\lambda$ parameter introduces a systematic error, we have to perform
a linear $\lambda \to 0$ extrapolation as shown in figure~\ref{fig:extra}. 
The extrapolated result $\beta=4.938(4)$ (including the phase factor) and the
result from multi-parameter re-weighting \cite{Fodor:2001au} are in very good
agreement. The $\lambda$ dependence is expected to be smaller for larger $\mu$,
therefore from now on we only give results for $\lambda/m=0.2$.
\FIGURE{
\label{fig:sus}
\includegraphics[height=7.5cm]{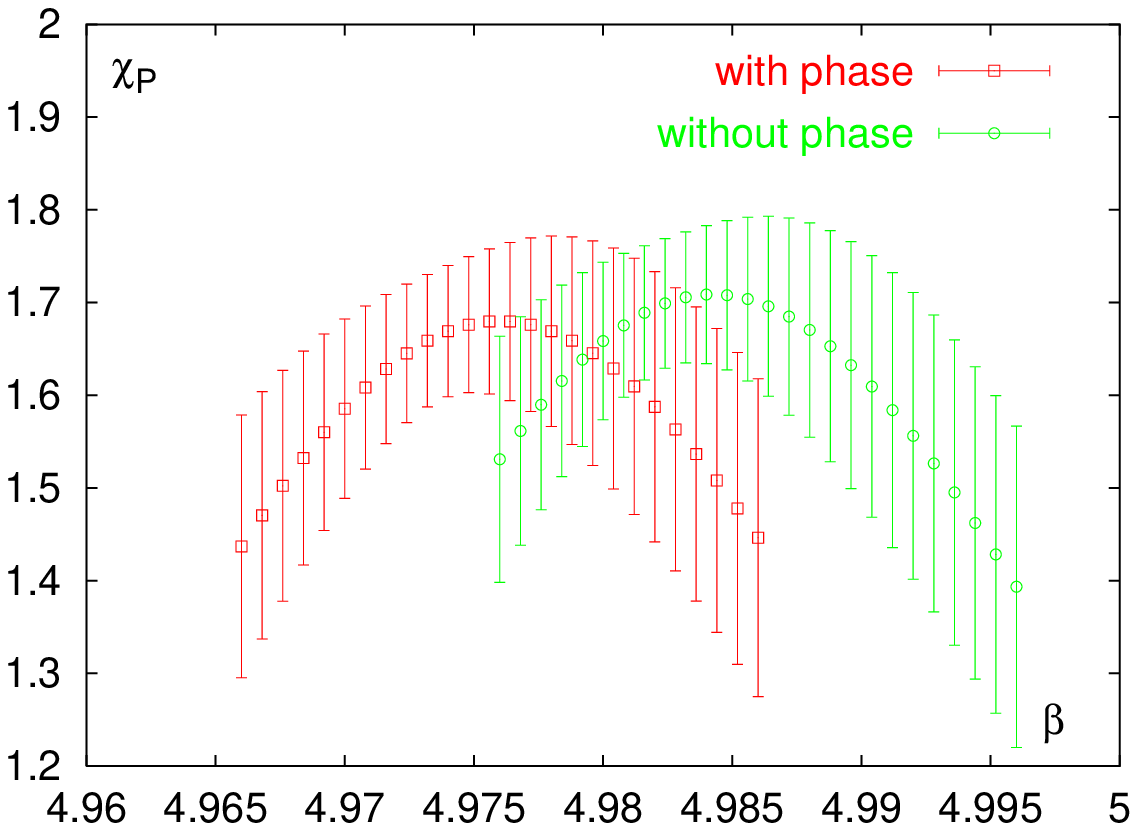}
\caption{Simulation parameters as in figure~1. Shown is the plaquette
  susceptibility as function of the coupling $\beta$. }
}  
\FIGURE{
\label{fig:extra}
\includegraphics[height=7.5cm]{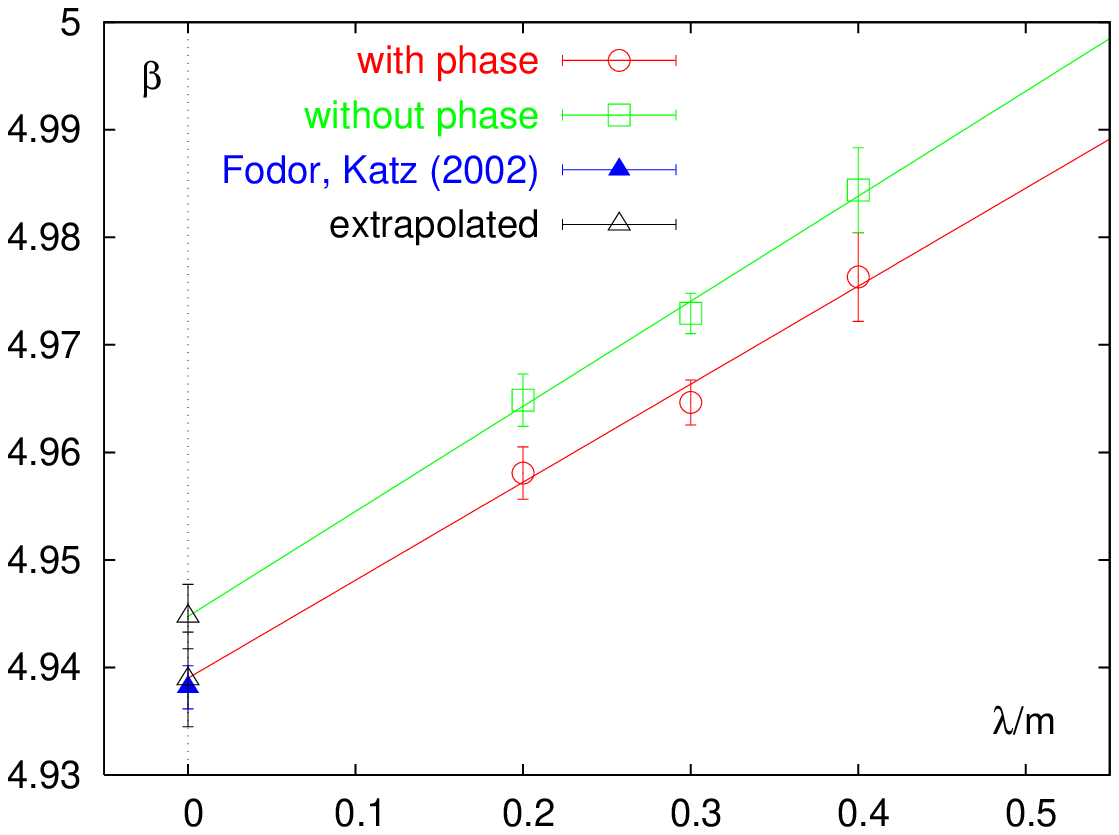}
\caption{The $\lambda\to 0$  extrapolation of the critical couplings on the
  $4^4$ lattice at $a\mu=0.3$. }
}
 
\section{The Plaquette expectation value and the phase diagram \label{sec:phases}}

Let us now discuss results for the plaquette expectation values from the $6^4$
lattice as shown in figure~\ref{fig:plaq1} and figure~\ref{fig:plaq2}.  
\FIGURE{
\label{fig:plaq1}
\includegraphics[height=7.5cm]{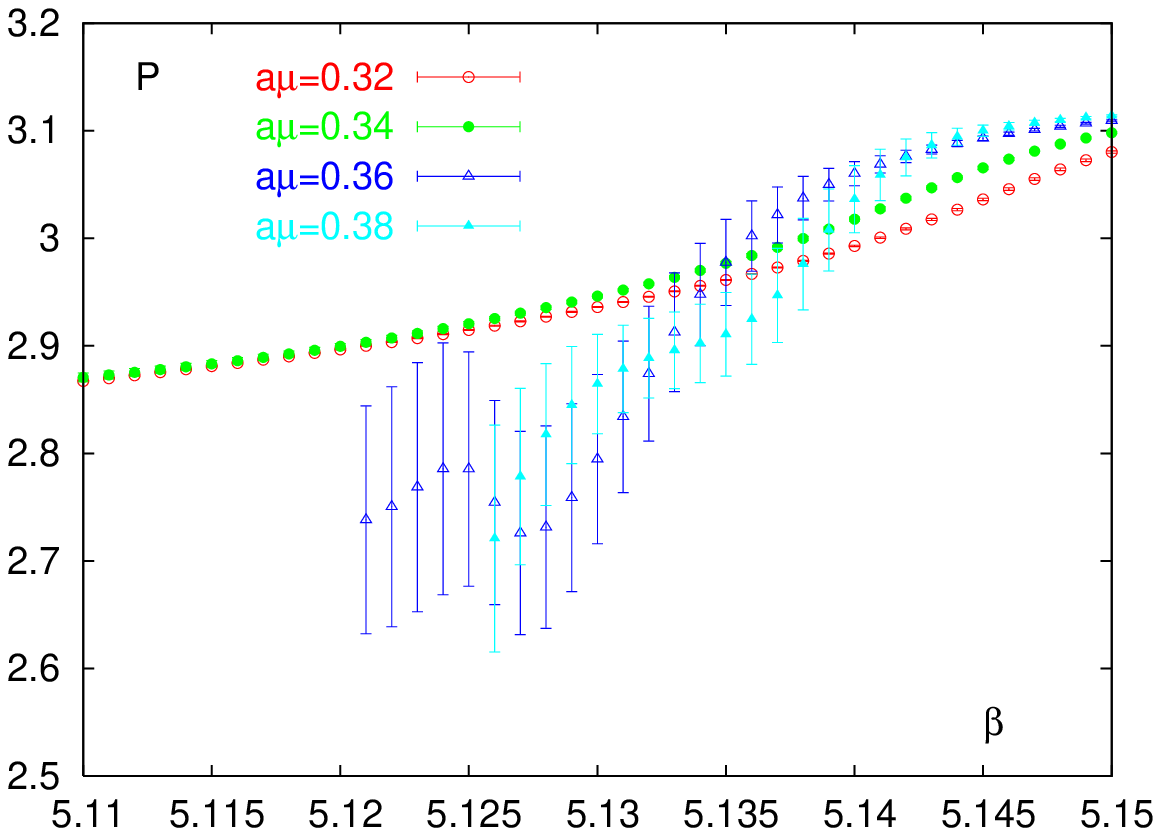}
\caption{Results for simulations at $\lambda=0.01$,
  $n_f=4$, $am=0.05$, and number of lattice points: $6^4$. Shown is the
  Plaquette expectation value as a function of the coupling $\beta$ for
  different chemical potentials }
}
\FIGURE{
\label{fig:plaq2}
\includegraphics[height=7.5cm]{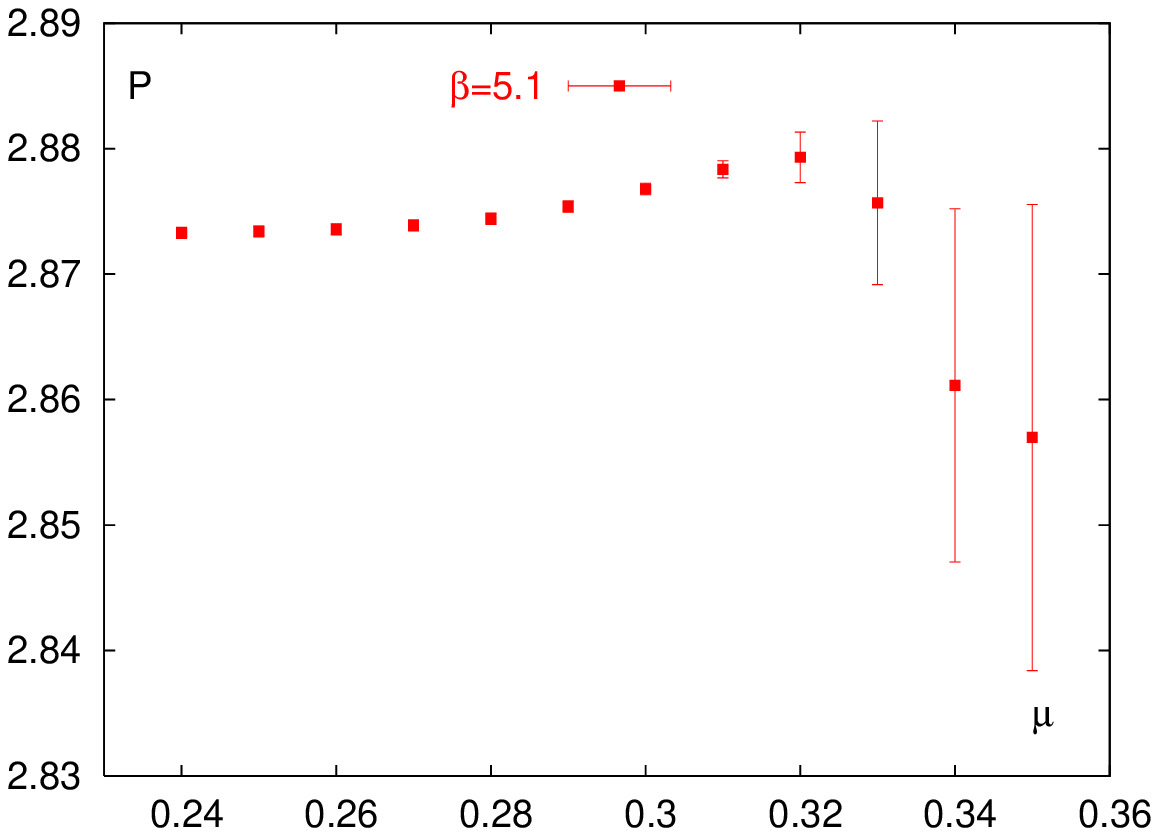}
\caption{Simulation parameters as in figure~5. Shown is the plaquette
  expectation value at fixed coupling, as a function of the chemical potential. }
}

At chemical potentials $\mu\lesssim 0.36$, the plaquette signals the QCD
transition through a rapid crossover from a low temperature phase of 
$<P>\sim 2.9$ 
to a high temperature phase of $<P>\sim 3.1$. For $\mu\gtrsim 0.36$ the plaquette
expectation value at small temperatures drops to $<P>\sim 2.85$. This new low
temperature phase of the plaquette at high   
chemical potentials is caused by the fermion determinant. As one can see in
figure~\ref{fig:det_phase} the region around $P\sim 2.85$ is the region
which is less suppressed by the phase factor. Another interesting observation
is that the critical coupling, which is decreasing in $\mu$ for $\mu<0.36$
starts to increase for $\mu>0.36$. 

The plaquette expectation value thus suggests
the existence of three different phases in the ($T$,$\mu$)-diagram with a
triple point, where all those phases coincide. In
figure~\ref{fig:phase_diagram} we show the phase diagram in physical units. 
\FIGURE[t]{
\label{fig:phase_diagram}
\includegraphics[height=7.5cm]{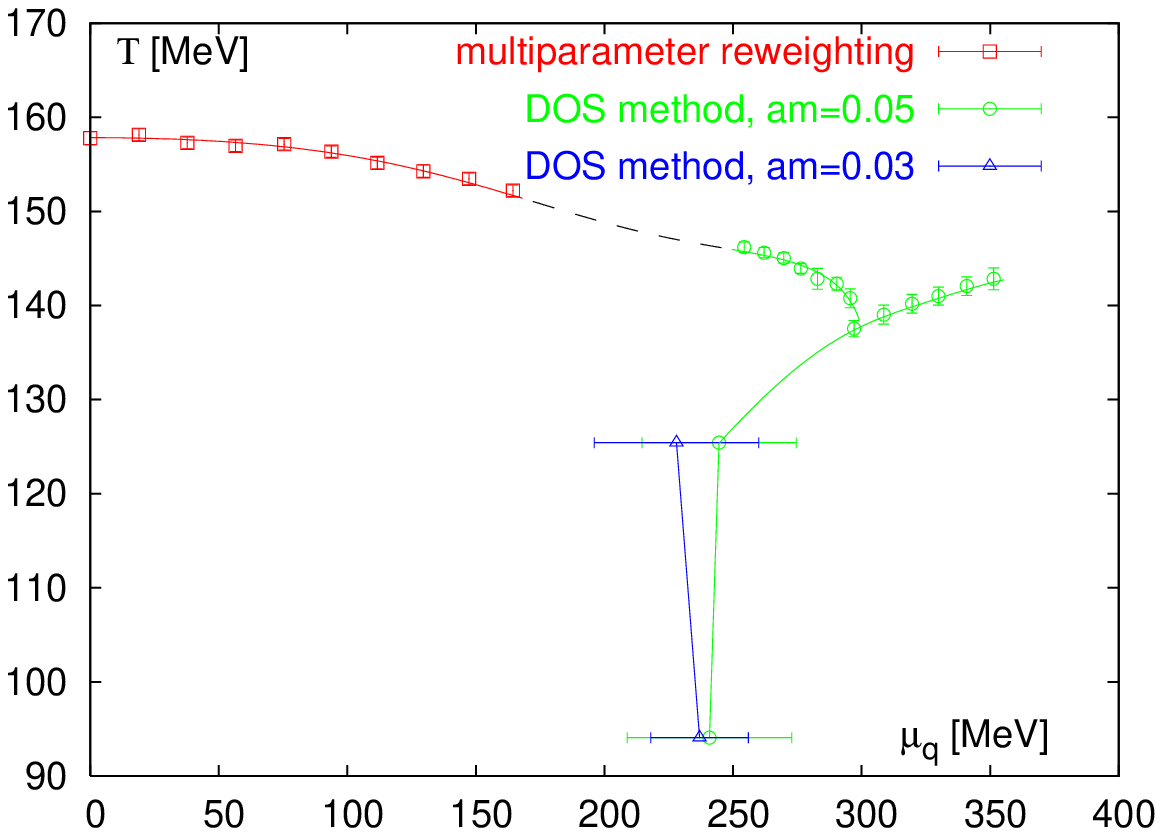}
\caption{The phase diagram in physical units from $N_t=6$ and 8 lattices. }
}
The phase boundaries were determined by calculating the peaks in the plaquette
susceptibility. The points at $T=93~\mbox{MeV}$ are from calculations on $6^3\times 8$
lattices. Note, that we make no statement about the order of the 
transition lines. To determine the order of the transition one has to perform
a finite-size-scaling analysis which is beyond the scope of this article.

The triple point is located around $\mu_q^{\rm tri}\approx 300~\mbox{MeV}$,
however its temperature ($T^{\rm tri}$) decreases from $T^{\rm tri}\approx
148~\mbox{MeV}$ on the $4^4$ lattice to $T^{\rm tri}\approx 137~\mbox{MeV}$ on
the $6^4$ lattice. This shift reflects the relatively large cut-off effects one
faces, with standard staggered fermions and temporal extents of 4 and 6. 

Also shown in figure~\ref{fig:phase_diagram} are points from
simulations with quark mass $am=0.03$. The phase boundary at low temperatures
turned out to be --- within our statistical uncertainties --- independent of
the mass. This gives evidence that the transition is associated with the
onset of baryonic matter rather than a pion condensate. Going from $am=0.05$
to $am=0.03$, the pion mass changes by a factor of $\approx\sqrt{3/5}=0.77$
whereas the nucleon mass remains approximately constant.

\section{The quark number density \label{sec:thermodynamics}}
To reveal the properties of the new phase located in the lower right corner of
the phase diagram, we calculated the quark number density, at constant coupling
$\beta$ and at constant temperature respectively. 
To obtain the density $n_q$
we perform the following integration
\begin{equation}
\label{eq:ddmu}
\left<\frac{{\rm d} \ln {\rm det}M}{{\rm d}(a\mu)}\right>
=\int dx\; \left<\frac{{\rm d} \ln {\rm det}M}{{\rm
      d}(a\mu)}\cos(\theta)\right>_x \rho(x)
\label{eq:dmusq}
\end{equation}
The thermodynamic quantity $n_q$ are given as usual by
\begin{equation}
n_q =
\frac{1}{a^3 N_s^3 N_t}
\left<\frac{{\rm d} \ln {\rm det}M}{{\rm d}(a\mu)}\right>
\end{equation}
In figure~\ref{fig:density} we show the baryon number density, which is related
to the quark number density by $n_B=n_q/3$.
\FIGURE[t]{
\label{fig:density}
\includegraphics[height=7.5cm, width=10.5cm]{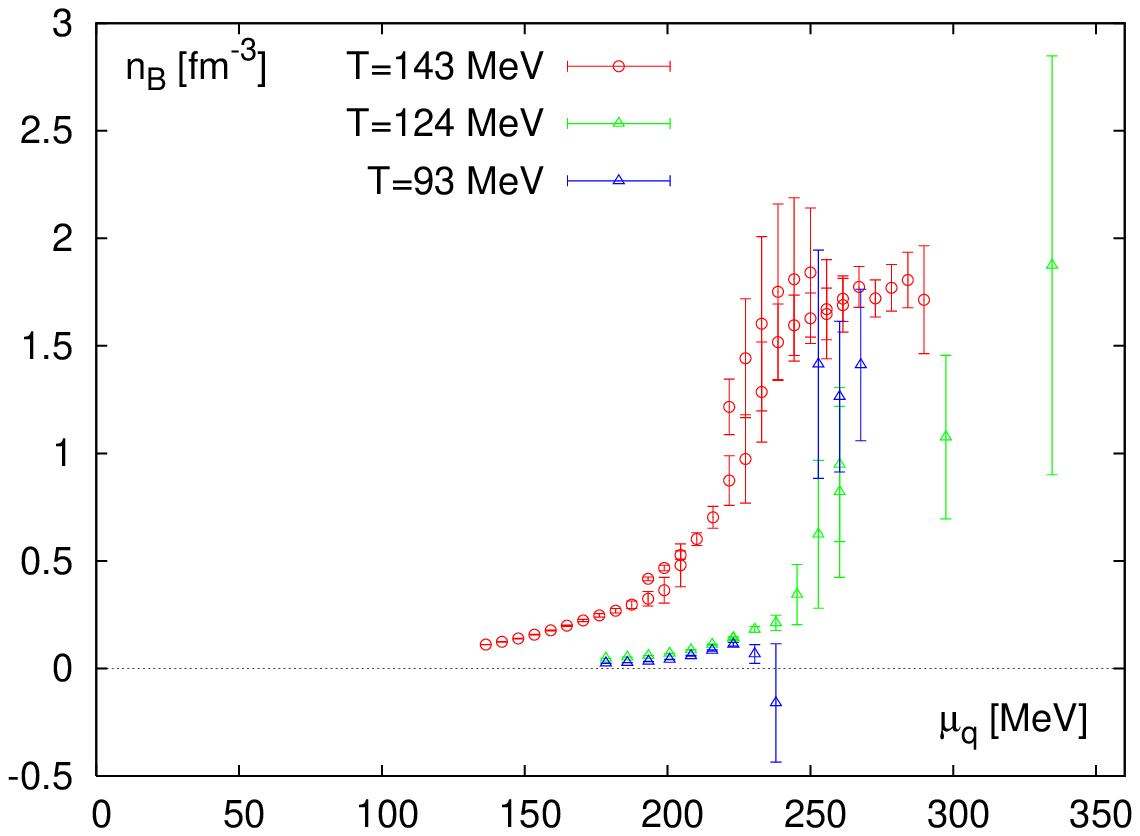}
\caption{The quark number density at constant temperature $T=143~\mbox{MeV}$
  ($4^4$ lattice), $T=124~\mbox{MeV}$ ($6^4$ lattice) and $T=93~\mbox{MeV}$
  ($6^3\times 8$ lattice).}
}
The results are plotted in
physical units and correspond to a constant temperature of $T\approx
143~\mbox{MeV}$ ($4^4$ lattice), $T\approx 124~\mbox{MeV}$ ($6^4$ lattice) and
$T\approx 93~\mbox{MeV}$ ($6^4\times 8$ lattice). In order to remove the leading
order cut-off effect,
we have multiplied the data with the factor $c=SB(N_t)/SB$, which
is the Stefan-Boltzmann value of a free lattice gas of quarks at a given value of $N_t$,
divided by its continuum Stefan-Boltzmann value. For unimproved staggered
fermions this correction factor can be as large as 2 and will not completely
remove all the cutoff effects. 

At the same value of the chemical potential where we find also a peak in the
susceptibility of the plaquette ($\mu_c)$, we see a sudden rise in the baryon
number density. Thus for $\mu>\mu_c$ we enter a phase of dense matter. The
transition occurs at a density of $(2-3)\times n_N$, where $n_N$ denotes
nuclear matter density. Above the transition, the density reaches values of
$(10-20)\times n_N$. Quite similar results have been obtained recently by
simulations in the canonical ensemble \cite{canonical}. 

\section{Conclusions \label{sec:conclusion}}
We have explored the QCD phase diagram by first principle lattice calculations
at so far unreached hight densities. In the accessible region
of $T\gtrsim 100~\mbox{MeV}$ and $\mu_q\lesssim 400~\mbox{MeV}$ we have been able to
identify three different regions, which seem to be separated by different
plaquette expectation values and quark number densities. These regions
coincide in a triple-point. The triple-point has been located at 
$T^{\rm tri}\lesssim 137~\mbox{MeV}$ and $m_q^{\rm tri} \approx 300~\mbox{MeV}$ at
finite lattice spacing ($N_t=6$). 
We note, that the lowest reachable temperature on $N_t=6$, is about $60~\mbox{MeV}$
(setting the scale e.g. by $m_{\rho}$). We thus find the triple-point about 
$80~\mbox{MeV}$ above the lowest temperature. This is 
the first numerical evidence from lattice QCD for a third phase (appart form the 
hadronic phase and the quark gluon plasma) and a triple
point in the QCD phase diagram. In the third phase the quark number density
reaches values of $(10-20)\times n_N$, where $n_N$ denotes normal nuclear
matter density. 

The new phase is a natural candidate for a color superconducting phase.
Recently, by combining experimental results from cold atoms in a trap \cite{atoms}
and some universal arguments, an upper bound for the transition line from the 
quark gluon plasma phase (QGP) to the superconducting phase (SC) was proposed 
($T_c\le0.35 E_F$) \cite{Schaefer, Shuryak}. To first approximation the Fermi-Energy 
$E_F$ is given by the chemical potential $\mu_q$. In \cite{Shuryak} the triple point was 
estimated by comparing this upper bound with the experimental freeze-out curve. 
A temperature of $T^{\rm tri}\le70~\mbox{MeV}$ was found. 
Our value of the triple-point roughtly corresponds to $T_c\le0.46 E_F$.
It is interesting that the two values are close.

At low temperatures we find a phase boundary which is very steep
and almost independent of $T$. Although our lowest temperature is 
$96~\mbox{MeV}$ an extrapolation to $T=0$ seems to be reasonable and would
yield a critical chemical potential of $\mu_q(T=0)\approx 250~\mbox{MeV}$ 
or equivalently $\mu_B/T_c(\mu_B=0)\approx 4.7$.
This number appears to be at the lower edge of the phenomenological expectation
of $\mu_B/T_c(\mu_B=0)\approx 5-10$. Note, that our lattice spacing is 
close to the strong coupling regime and we should feel the influence of the
strong coupling limit. Strong coupling expansion calculations in general yield
much lower values of $\mu_B/T_c(\mu_B=0)\lesssim 1.5$
\cite{Kawamoto:2005mq}.

For this work the density of state method has been employed, which works 
well on small lattices up to chemical potentials of $\mu_q/T\lesssim 3$ (other
methods \cite{Fodor:2001au, methods} worked up to $\mu_q/T\lesssim 1$). The
method is however extremely expensive and thus will in the near future not
yield results close to the thermodynamic limit or the continuum limit, due
to limitations in computer resources. 

We have to emphasize once more, that the simulations have been carried out on
coarse lattices with an unphysical value of $n_f=4$ degenerate fermion flavor,
and that neither the continuum nor the thermodynamic limit has been
taken. Since we used unimproved staggered fermions, the corrections due to a 
finite lattice spacing are large. We also expect corrections due to the finite 
size of our volume. The simulations have not been performed with
a constant quark mass, but $m_q/T=0.3$ has been held fixed.  

\section*{Acknowledgments} 

CS would like to thank F. Karsch, Ph.~de~Forcrand and K.~Splittorff for
helpful discussions and comments. CS has partially been supported by a
contract DE-AC02-98CH1-886 with the U.S. Department of Energy. ZF and SK have
been supported by OTKA Hungarian Science Grants No.\ T34980, T37615, M37071,
T032501, AT049652 and the German Research Grand (DFG) FO 502/1. The
computations were carried out at E\"otv\"os University on the 330 processor PC
cluster of the Institute for Theoretical Physics and the 1k node PC cluster
ALiCEnext at the University of Wuppertal, using a modified version of
the publicly available MILC code \cite{MilcCode} and a next-neighbor
communication architecture \cite{Fodor:2002zi}.  


\end{document}